\documentclass[twocolumn,floatfix]{revtex4}
\usepackage{times}
\usepackage{amsmath}
\usepackage{graphicx}

\newcommand{\bfh}{\ensuremath{\boldsymbol{h}}}

\begin{document}
\sloppy

\title
{Charge-Transfer Matrix Elements by FMO-LCMO Approach:
Hole Transfer in DNA 
with Parameter Tuned Range-Separated DFT}

\author{Hirotaka Kitoh-Nishioka\footnote{
Present address: Department of Chemistry, Graduate School of Science, Nagoya University,
Furo-cho, Chikusa-ku, Nagoya 464-8601, Japan.
E-mail: kito.hirotaka@b.mbox.nagoya-u.ac.jp} 
and Koji Ando\footnote{
E-mail: ando@kuchem.kyoto-u.ac.jp
}}

\affiliation{
Department of Chemistry, Graduate School of Science, Kyoto University, 
Sakyo-ku, Kyoto 606-8502, Japan}

\begin{abstract}
A scheme for computing charge-transfer matrix elements with the linear combination of fragment molecular orbitals and the \lq nonempirically tuned range-separated\rq\ density functional is presented. It takes account of the self-consistent orbital relaxation induced by environmental Coulomb field and the exchange interaction in fragment pairs at low computational scaling along the system size. The accuracy was confirmed numerically on benchmark systems of imidazole and furane homo-dimer cations. Applications to hole transfers in DNA nucleobase pairs and in a $\pi$-stack adenine octomer highlight the effects of orbital relaxation.
\end{abstract}

\maketitle

\section{Introduction}

Charge transfer (CT) reactions are of fundamental and broad importance 
in chemistry, biochemistry, and materials science \cite{Marcus1985,Bixon1999}.
The CT rate and mechanism are determined by diagonal site energies and off-diagonal transfer integrals
of the Hamiltonian matrix,
which are thus key to elucidating and potentially controlling the reaction \cite{Coropceanu2007,Grozema2008}.

For quantitative evaluation of the Hamiltonian matrix elements, 
the effects of orbital relaxation and electron correlation can be significant.
The state-of-the-art wave function approaches such as 
the multi-reference configuration-interaction 
and 
the coupled-cluster
methods
are still computationally expensive \cite{Pieniazek07}, especially for combining with
molecular dynamics simulations \cite{Oberhofer12}.
In this regard, the recently emerging long-range corrected (LC) density functional theory (DFT)
is attractive, 
as it takes an improved account of the CT interaction by reducing the self-interaction error
in DFT while maintaining the short-range dynamic correlation effects \cite{Iikura2001,Tawada2004}.
This aspect 
on the redox site energies
has been examined recently by
a \lq nonempirically tuned range-separated (NET RS)\rq\ DFT \cite{Foster2012}, 
in which the range-separation parameter 
for decomposing the two-electron Coulomb interaction 
is optimized in accord with Janak's theorem 
to reproduce 
the ionization potential (IP) and electron affinity 
by the orbital energies of highest-occupied (HO) and lowest-unoccupied 
Kohn-Sham (KS) 
molecular orbitals (MOs) \cite{Stein2009}.
The optimal parameters for the nucleobases of DNA and RNA have been thus reported \cite{Foster2012}.

However, 
the parameters optimal for the intramolecular site energies are
not necessarily equally appropriate for
the off-diagonal intermolecular transfer integrals.
Nonetheless,
we know empirically that the off-diagonal energies are approximated by the
average of diagonal energies multiplied by the overlap integral and an empirical factor,
as known prototypically for the extended H{\"u}ckel method.
In addition, 
the one-electron mean-field picture has been considered appropriate
for the transfer integrals
\cite{Newton80,Yang06,Nishioka11a}
due to cancelation of 
the dynamic electron-correlation effect, as the transfer integrals
correspond to the difference between 
the energies of two adiabatic electronic states at the diabatic surface crossing.
These aspects are thus intriguing for quantitative evaluation.

The method for computing the off-diagonal transfer integrals is not unique:
it depends on the definition of basis states.
This is analogous to the arbitrariness in defining the diabatic states 
for non-adiabatic transitions \cite{Pacher88,Ando94,Subotnik2008}.
When the redox sites or moieties are spatially well-defined, 
as is the case for the $\pi$-stack DNA nucleobases,
the MOs of isolated fragments provide a reasonable choice,
as employed in the conventional fragment-orbital approaches \cite{Troisi2001,Senthilkumar2005,Kubar2008}.
However, 
the MOs optimized for isolated fragments lack the orbital relaxation
induced by the long-range Coulomb and short-range exchange interactions.
If we are to include these interactions, we need to handle large systems
and have to determine the procedure for transforming 
the resulting delocalized canonical MOs to localized diabatic bases.
Thus, there exists a practical trade-off 
between the inclusion of orbital relaxation
and the diabatization.
The methods of constrained DFT \cite{QinWu06,delaLande10} 
and frozen density embedding \cite{Pavanello13}
are recently proposed prescriptions to this problem. 

In this work we put forward another solution by employing
the method of fragment molecular orbitals (FMO) \cite{Kitaura1999, Nakano2002, Fedorov2007,Tanaka2014} 
and their linear combinations (FMO-LCMO) \cite{Tsuneyuki2009,Kobori2013,Nishioka2011,Kitoh-Nishioka2012}.
The FMO-LCMO method was originally proposed to obtain canonical MOs of large systems \cite{Tsuneyuki2009,Kobori2013}.
The FMO method first decomposes the total system into fragments
and optimizes the MOs of each fragment self-consistently under the Coulomb field of other fragments.
Then, dimer, trimer, or tetramer calculations are carried out to take account of the exchange interactions.
In the FMO-LCMO method, the diagonal elements of Hamiltonian matrix of the total system
are computed from
the results of these fragment calculations in a form to remove the excess counting of energy,
whereas the off-diagonal elements are constructed from the dimer or trimer matrices projected to the monomer FMO space.
The total Hamiltonian matrix 
is then diagonalized 
to find the canonical MOs of the entire system.
Now, it is seen that the Hamiltonian matrix before the diagonalization provides
appropriate diagonal site energies and off-diagonal transfer integrals 
for the study of CT reactions.

After formulating this idea in Section \ref{sec:method},
Section \ref{sec:results} presents 
an assessment of the scheme on benchmark systems and
numerical applications to $\pi$-stack DNA nucleobases. 
Section \ref{sec:conclusion} summarizes and concludes. 

\section{Method}
\label{sec:method}

Here we outline the FMO-LCMO method \cite{Tsuneyuki2009,Kobori2013,Nishioka2011,Kitoh-Nishioka2012}.
For simplicity, we restrict to the FMO2 version with dimer exchange,
although extension to the trimer FMO3 is straightforward \cite{Kobori2013}.
In the FMO2 method, the orbitals of each fragment are optimized
self-consistently under the Coulomb field of other fragments.
The resultant $p$th orbital of fragment $I$ is denoted by
$\varphi_{p}^{I}$.
The fragment dimer calculations are then carried out under the
Coulomb field of other monomer fragments determined above.
In the FMO-LCMO method, the intra- and inter-fragment Hamiltonian matrix elements
are defined by
\begin{align}
H^{\rm (total)}_{Ip,Iq} &= \sum_{J \ne I}
\langle \varphi_p^I | \bfh_{}^{IJ} | \varphi_q^I \rangle
- (N-2) 
\langle \varphi_p^I | \bfh_{}^{I} | \varphi_q^I \rangle
\label{eq:HtotIpIq}
\\
H^{\rm (total)}_{Ip,Jq} &= 
\langle \varphi_p^I | \bfh_{}^{IJ} | \varphi_q^J \rangle
\hspace*{1.5em}
(I \ne J), 
\label{eq:HtotIpJq}
\end{align}
in which
$\bfh_{}^{I}$ and
$\bfh_{}^{IJ}$ 
are the KS (or Fock) matrices of fragment $I$ and fragment dimer $IJ$, respectively.
In Eqs. (\ref{eq:HtotIpIq}) and (\ref{eq:HtotIpJq}),
the notation $\langle \varphi_p^I | \bfh_{}^{IJ} | \varphi_q^J \rangle$
represents the dimer KS matrix projected to the monomer orbital space.
The diagonalization of this $H^{\rm (total)}$
has been demonstrated to give
accurate approximations to the canonical MOs and MO energies of 
large systems \cite{Tsuneyuki2009,Kobori2013,Nishioka2011,Kitoh-Nishioka2012}.
 
The proposal in this work is to employ
the matrix elements of $H^{\rm (total)}$ 
for the orbital site energies $\varepsilon_{Ip}$
and the transfer integrals $T_{Ip,Jq}$ of CT systems.
As seen from the procedure described above,
they include the orbital relaxation from the self-consistent Coulomb field 
and the pairwise exchange interactions.
This can be an advantage over methods that employ orbitals optimized for isolated fragments.

Because the orbitals of different fragments are optimized separately in the FMO method,
they are in general non-orthogonal
with finite overlaps
$S_{Ip,Jq} \equiv \langle \varphi_p^I | \varphi_q^J \rangle \ne 0$ for $I \ne J$.
Thus, the transfer integrals appropriate for description of CT processes are
\begin{align}
T_{Ip,Jq}' = 
(
H^{\rm (total)}_{Ip,Jq} 
- S_{Ip,Jq} ( H^{\rm (total)}_{Ip,Ip} 
 + H^{\rm (total)} _{Jq,Jq} ) / 2   
)
\nonumber \\
/(1 - S_{Ip,Jq}^2)
\nonumber \\
 = (
T_{Ip,Jq} - S_{Ip,Jq} \left( \varepsilon_{Ip}
 + \varepsilon_{Jq} ) / 2
\right)/(1 - S_{Ip,Jq}^2)
, 
\label{eq:Tprime}
\end{align}
from symmetric-orthogonalized orbitals \cite{Lowdin1950}.

In what follows, we will compare the FMO-LCMO with the 
fragment-orbital approach (FOA) \cite{Troisi2001,Senthilkumar2005,Kubar2008} 
and the generalized Mulliken-Hush (GMH) method \cite{Cave97}.
Both FOA and GMH
have been often employed and well documented in the literature \cite{Kitoh-Nishioka2012,Felix2011,Blancafort2006,Valeev2006}.
However, we outline them below for convenience.

In the FOA, 
the orbitals $\phi_i$ are first optimized for isolated monomers.
Next, 
the KS matrix of aimed complex
(a dimer or larger assembly of fragment monomers),
$\bfh_{\rm KS}^{\rm (cmplx)}$, is constructed
with or without orbital optimization.
The site energies $\varepsilon_i$ and transfer integrals $T_{ij}$ are then
computed with
\begin{align}
\varepsilon_i &= 
\langle \phi_i | \bfh_{\rm KS}^{\rm (cmplx)} | \phi_i \rangle
\label{eq:FOAsiteE}
\\
T_{ij} &= 
\langle \phi_i | \bfh_{\rm KS}^{\rm (cmplx)} | \phi_j \rangle
\label{eq:FOATij}
\end{align}
Because the orbitals of different fragments are not orthogonal,
similarly to the case of FMO-LCMO noted above, 
the transfer integrals $T'_{ij}$ are computed with the formula
equivalent to Eq. (\ref{eq:Tprime}). 

In the GMH method for hole transfers with a one-electron KS-MO approach \cite{Felix2011}, 
the transfer integral $T'_{ij}$ 
is approximated by the HOMO (H) and HOMO-1 (M) energy difference for the
neutral dimer fragment of sites $i$ and $j$,
scaled by the dipole moment matrix elements with respect to the two MOs as
\begin{equation}
T'_{ij} = 
\frac{( \varepsilon_{\rm H}^{\rm (dim)} - 
 \varepsilon_{\rm M}^{\rm (dim)} )
| \mu_{\rm H, M} |}
{\sqrt{(\mu_{\rm H, H} - \mu_{\rm M, M} )^2 + 
4 ( \mu_{\rm H, M} )^2}},
\end{equation}
\begin{align}
\mu_{\rm H, M} &=
-\sum_{\alpha \beta } C_{\alpha, \rm H}^{\rm (dim)} 
C_{\beta, \rm M}^{\rm (dim)} 
d_{\alpha \beta}, 
\\
\mu_{\rm H, H} - \mu_{\rm M, M} &=
\sum_{\alpha \beta } (  C_{\alpha, \rm H}^{\rm (dim)} 
C_{\beta, \rm H}^{\rm (dim)} -
C_{\alpha, \rm M}^{\rm (dim)} C_{\beta, \rm M}^{\rm (dim)} 
)
d_{\alpha \beta}. 
\end{align}
Here, $C_{\alpha, \rm H}^{\rm (dim)}$ 
and $C_{\alpha, \rm M}^{\rm (dim)}$ are 
the MO coefficients for the dimer fragment
in the atomic orbital (AO) representation and 
$d_{\alpha \beta}$ are the dipole matrix element between
the AOs $\alpha$ and $\beta$.

All electronic calculations in this Letter were performed 
on neutral monomers, dimers, and octomers using the GAMESS program \cite{Schmidt1993}.

\begin{figure*}
\begin{center}
\includegraphics[width=0.35\textwidth]{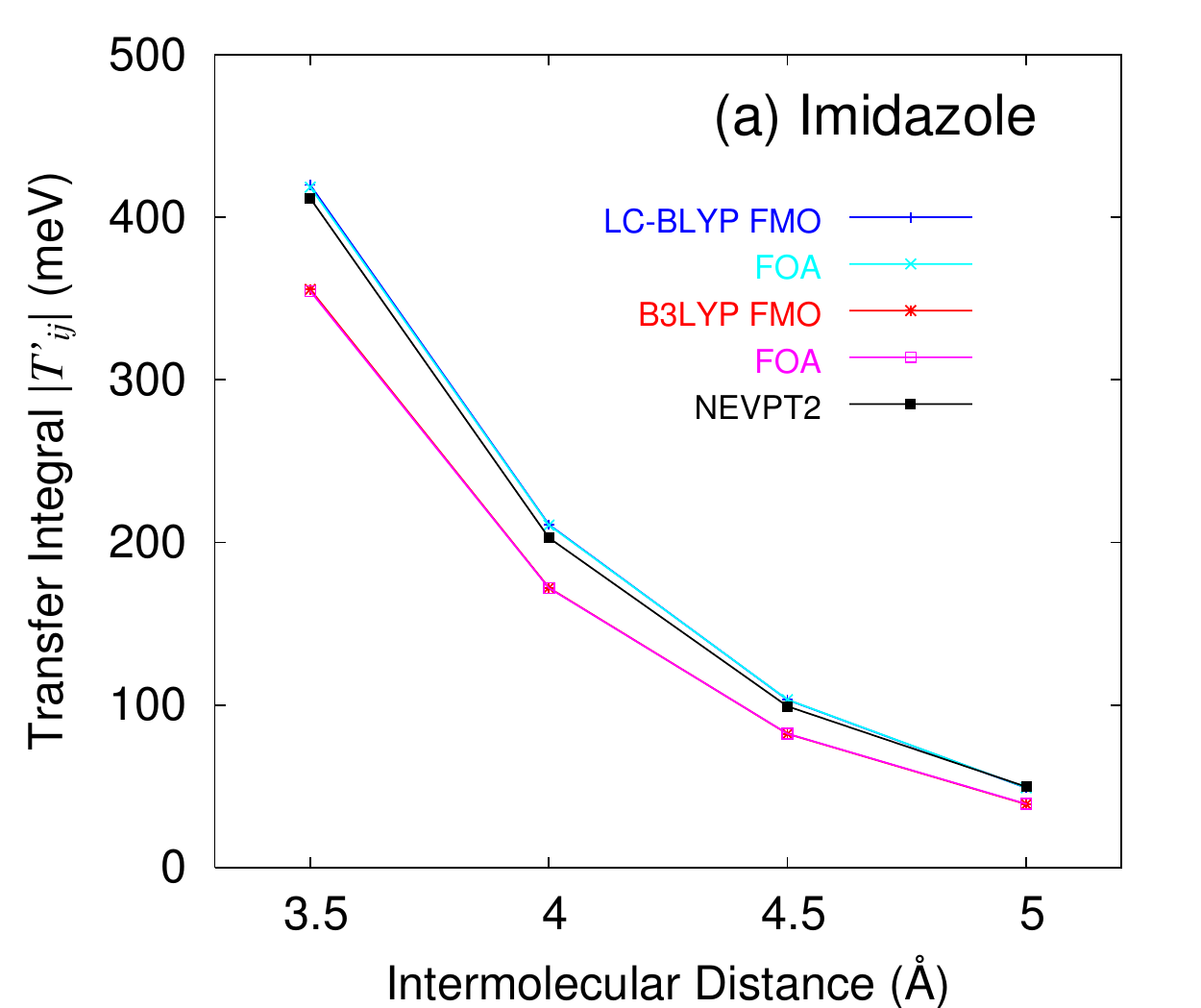}
\includegraphics[width=0.35\textwidth]{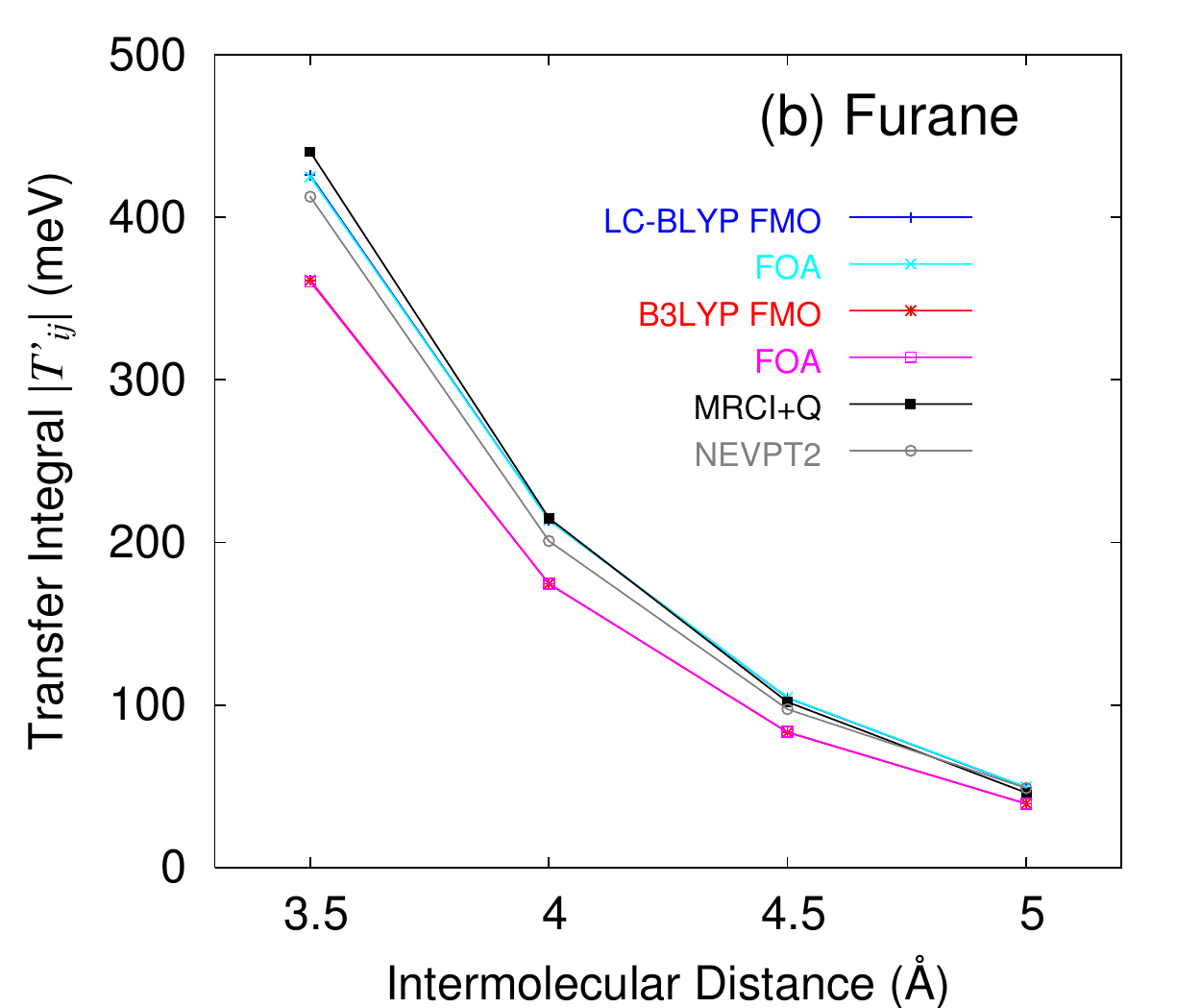}
\end{center}
\caption{
Transfer integrals for (a) imidazole and (b) furane homo-dimer cations
compared with the high-level \textit{ab initio} references \cite{Kubas2014}.
In both (a) and (b), FMO-LCMO and FOA with the same functional almost completely overlap.
}
\end{figure*}

\section{Results and discussion}
\label{sec:results}

\subsection{Assessment on benchmarks: imidazole and furane homo-dimer cations}

We first assess the
nonempirically tuned range-separated (NET RS) functional
for transfer integrals.
The reference data are taken from
a benchmark database \cite{Kubas2014} 
with high-level \textit{ab initio} calculations
of multireference configuration interaction (MRCI+Q) and
$n$-electron valence state perturbation theory (NEVPT2).
From the database we have chosen imidazole and furane homo-dimer cations
because their molecular structures are most related to the nucleobases.

The optimal values of range-separation parameter $\mu$ 
in the formula
\(
{1}/{r_{12}} = (1 - {\rm erf} ( \mu r_{12}))/{r_{12}} + {\rm erf} ( \mu r_{12})/{r_{12}}
\)
with the LC-Becke\cite{Becke1988}-Lee-Yang-Parr\cite{Lee1988} (LC-BLYP) functional \cite{Tawada2004}
for imidazole and furane monomers
were searched
with the procedure described in Ref. \cite{Foster2012},
and were found to be $\mu = 0.33$ bohr$^{-1}$ for both molecules.
Here we used the cc-pVTZ basis set for heavy atoms and cc-pVDZ for hydrogens,
although Ref. \cite{Kubas2014} employed
aug-cc-pVTZ for the former.
However, the results displayed in Figure 1 present high accuracy.
By contrast, the standard B3LYP functional \cite{Becke1993}
systematically underestimates the transfer integrals
for both systems.
The figure also indicate that the results from FMO-LCMO and FOA 
with the same functional are almost identical.
The numerical values are listed in Table S1 of the Supplementary materials.

\begin{figure}
\begin{center}
\includegraphics[width=0.35\textwidth]{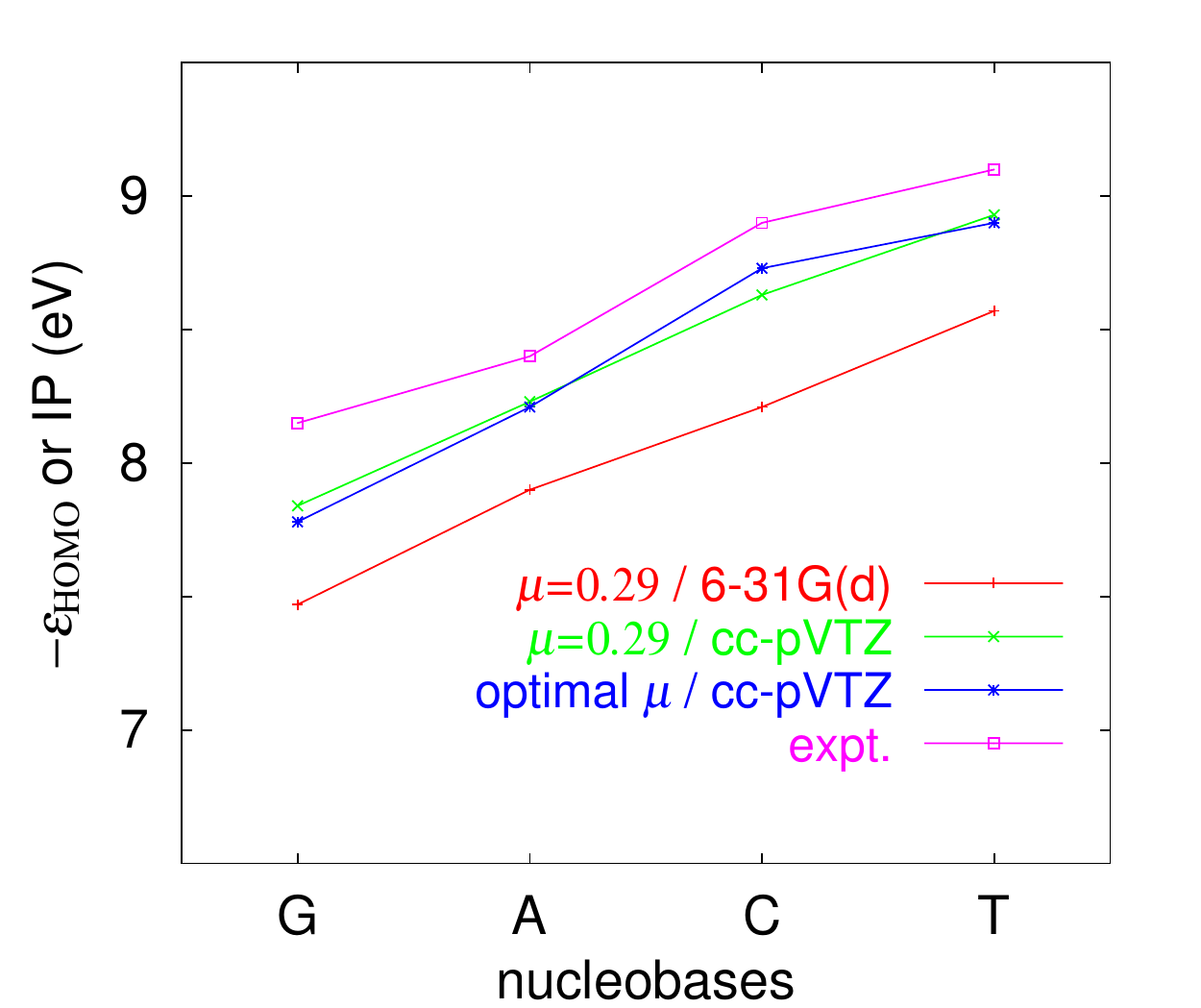}
\end{center}
\caption{
HOMO energies with LC-BLYP functional and experimental IPs 
of four nucleobases, guanine (G), adenine (A), cytosine (C), and thymine (T).
Each experimental IP has an uncertainty of $\pm 0.1$ eV.
}
\end{figure}

\begin{figure*}
\begin{center}
\includegraphics[width=0.35\textwidth]{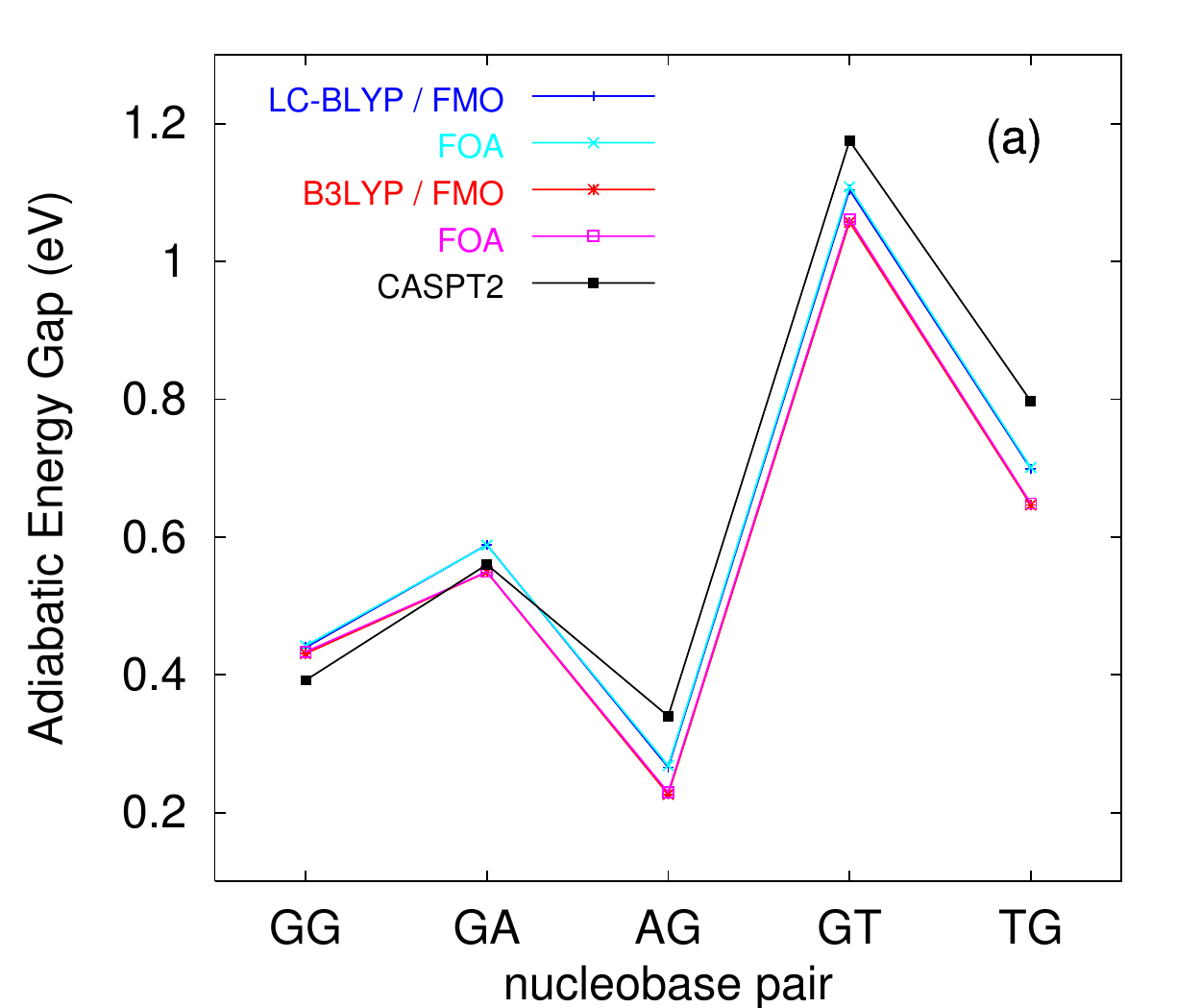}
\includegraphics[width=0.35\textwidth]{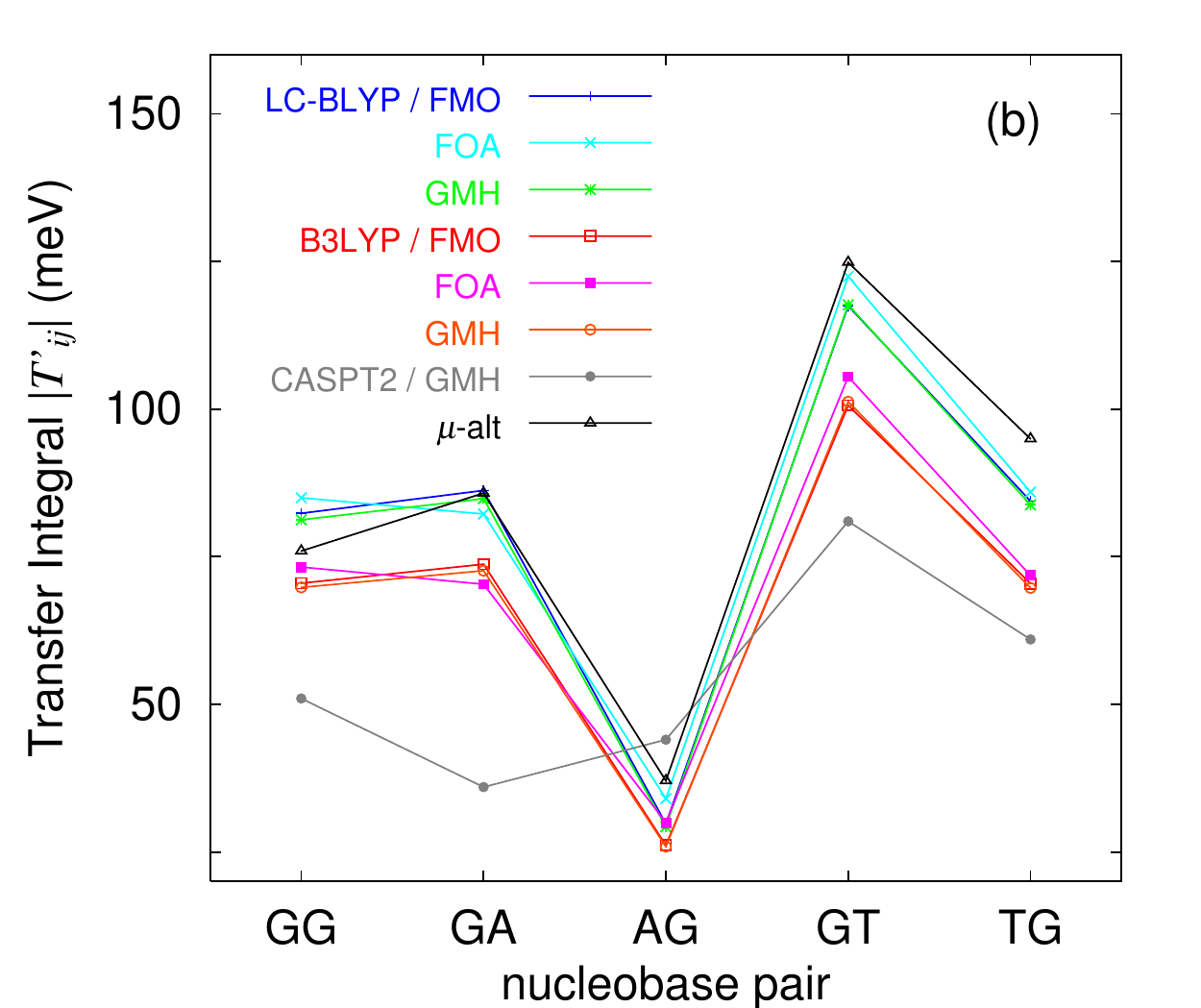}
\end{center}
\caption{
Adiabatic energy gaps $\Delta E'_{ij}$ (a) and hole-transfer integral $|T'_{ij}|$ (b) 
of nucleobase pairs GG, GA, AG, GT, TG,
with 6-31G(d) basis-set.
In (b), \lq $\mu$-alt\rq\ refers to a GMH calculation with the energy gap from CASPT2 
and the dipole moment matrix elements from LC-BLYP.
In (a), FMO-LCMO and FOA with the same functional almost completely overlap.
In (b), 
FMO-LCMO and GMH overlap for AG, GT, and TG
for both functionals. 
}
\end{figure*}

\subsection{Isolated nucleobases with a single range-separation parameter}

We now proceed to the nucleobase molecules.
The optimal values of the parameter $\mu$ 
with the LC-BLYP functional
and cc-pVTZ basis set have been reported (in bohr$^{-1}$) as
0.2738 for guanine (G),
0.2853 for adenine (A),
0.2948 for cytosine (C),
and
0.2850 for thymine (T) \cite{Foster2012}. 
These were optimized separately for each nucleobase.
However, 
we need to determine a single value for the computation of intermolecular transfer integrals,
for which we have chosen $\mu = 0.29$
because the latter three nucleobases have approximately this value.
With this we then reexamined the site energies.
The nuclear coordinates are taken from Ref. \cite{Foster2012}.

The resultant HOMO-1 and HOMO energies 
are displayed in Figure 2.
They are compared with those from the original optimal $\mu$ for each nucleobase and
with the corresponding experimental IPs \cite{Faber2011}.
The numerical values are listed in Table S2 of the Supplementary materials.
We see that the parameter $\mu = 0.29$ 
yields the orbital energies sufficiently close to those 
with the optimal $\mu$ for each nucleobase. 
We have also carried out calculations with 6-31G(d) basis set
in an aim to compare with the previous works \cite{Felix2011, Blancafort2006},
which will be discussed in Section \ref{subsec:basepair}.
Although the 6-31G(d) basis set systematically underestimate the absolute values,
the relative HOMO energies among the nucleobases,
the most relevant quantities for the hole transfers,
indicated by the slope of the plots
are similar among the methods and the experiment.
Having in scope the calculations of larger systems, it is desired to keep the size
of basis-set as small as possible.
We thus employ the 6-31G(d) basis-set 
hereafter, until we examine the basis-set dependence in Section \ref{subsec:basisEffect}.

\subsection{Hole transfer integrals for nucleobase pairs}
\label{subsec:basepair}

Here we examine the CT energies for nucleobase pairs in a $\pi$-stack configuration.
As guanine is the key donor with the lowest IP, we study
GG, GA, AG, GT, and TG pairs.
The nuclear coordinates are taken from Ref. \cite{Blancafort2006}.
Because the reference data from CASPT2 calculation \cite{Blancafort2006} 
corresponds to 
the adiabatic energy gap
rather than the diabatic site energy difference,
we have also calculated the adiabatic energy gaps from the diabatic site energies and
the transfer integrals.
The numerical values are listed in Table S3 of the Supplementary materials.

The resultant adiabatic energy gaps and transfer integrals are
plotted in Figure 3
for the LC-BLYP functional with $\mu = 0.29$ compared to B3LYP 
and CASPT2.
Figure 3a indicates that 
the adiabatic energy gaps from LC-BLYP are closer to CASPT2.
We also see that the values from FMO-LCMO and FOA with the same functional coincide 
well to be indistinguishable in the figure;
the difference is approximately 1 meV as seen in Table S3.
The difference between FMO-LCMO and FOA is 
in the treatment of orbital relaxation of the basis MOs.
It is included in the former while not in the latter.
This effect is more apparent in the transfer integrals discussed next.

Figure 3b compares the transfer integrals $T'$.
For both LC-BLYP and B3LYP functionals, 
FMO-LCMO and GMH coincide well 
for AG, GT, and TG. 
Notably, their deviation from FOA is now apparent.
This is due to the effect of orbital relaxation 
that is included in GMH and FMO-LCMO but not in FOA.
The numerical values of the transfer integrals are listed in Table S4 of the Supplementary materials.

Figure 3b also indicates that the transfer integrals from CASPT2 notably deviate
from both LC-BLYP and B3LYP.
Considering the accuracy of LC-BLYP confirmed in Figure 1
and the limited active orbital space 
employed in the CASPT2 \cite{Blancafort2006},
we have tested a GMH calculation with the energy gap from CASPT2 
and the dipole moment matrix elements from LC-BLYP.
The results denoted by \lq $\mu$-alt\rq\ in Figure 3b
are now close to LC-BLYP.
It would be plausible that the limited CASPT2 calculations 
yield more reliable energy gaps than dipole matrix elements.
We also note in Figure 3b that B3LYP underestimates the transfer integrals for
all the nucleobase pairs, the same tendency as was observed in Figure 1.

\begin{figure}
\begin{center}
\includegraphics[height=0.3\textheight]{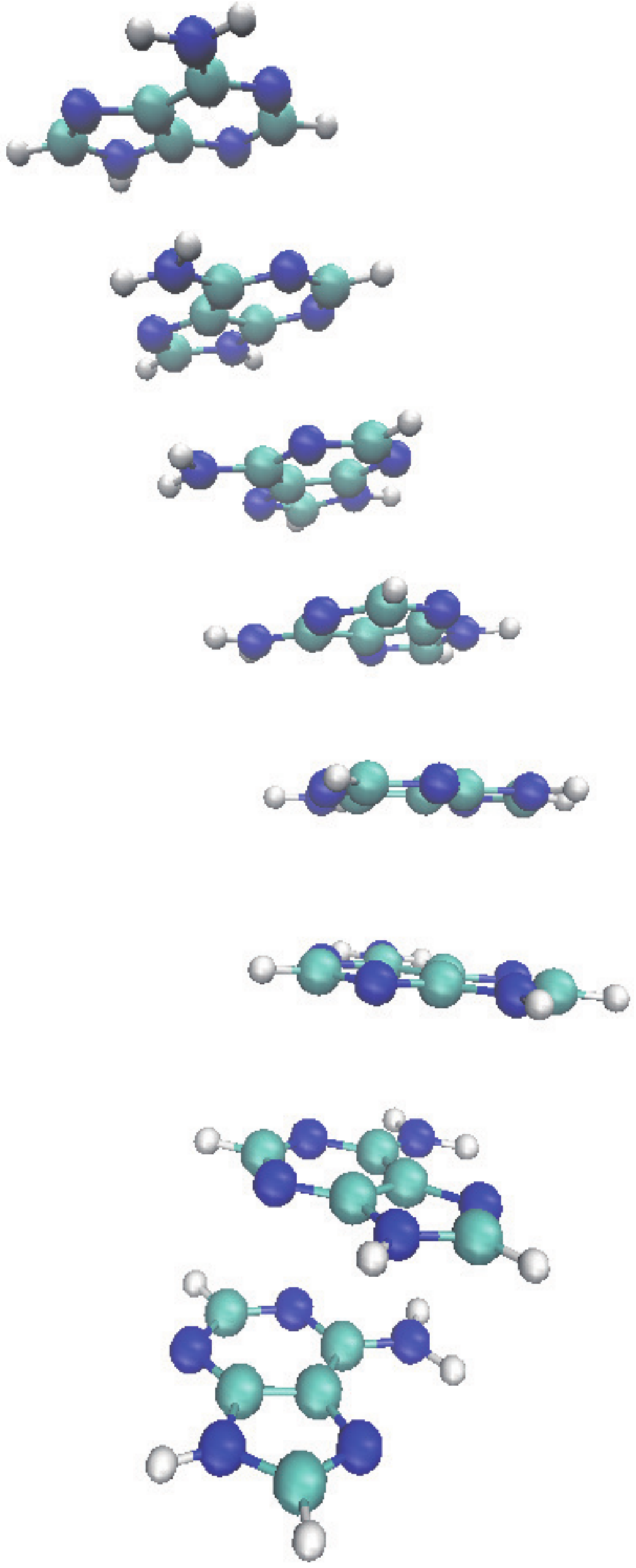}
\end{center}
\caption{
Structure of adenine octomer in the 
idealized B-DNA $\pi$-stack configuration.
}
\end{figure}

\subsection{$\pi$-Stack adenine molecules}
\label{subsec:adenines}

Next, we study a larger system, an adenine octomer in the 
idealized B-DNA $\pi$-stack configuration (Figure 4).
The structure of atoms other than hydrogen are constructed
by using the Web-3DNA software \cite{Zheng2009}.
The positions of hydrogen atoms are optimized for each isolated adenine
by B3LYP/6-31G(d) calculation.

As discussed above, 
one major difference between FOA and FMO-LCMO is in the treatment of orbital relaxation.
Another difference is in the scaling of computational cost along the system size.
In the standard use of FOA, 
the basis MOs are computed for each isolated monomer of donor and acceptor molecules.
Then, the KS matrix of the entire donor-acceptor system is constructed,
over which the matrix elements on the basis MOs from the monomer calculation 
are computed (see Eqs. (\ref{eq:FOAsiteE}) and (\ref{eq:FOATij})).
Therefore, to study the system of Figure 4, for instance, 
the full KS matrix 
$ \bfh_{\rm KS}^{\rm (total)}$ of the entire eight molecules should be calculated.
This 
will quickly become
computationally prohibitive as the system size increases.

An evasion of
this size-scaling problem in FOA 
would be to reduce the size of KS matrix to the dimer 
$ \bfh_{\rm KS}^{\rm (dim)}$ 
for each donor-acceptor pair.
In such cases, 
in order to take account of the orbital relaxation, the calculations may be performed
under the classical electrostatic, or the \lq molecular-mechanical (MM)\rq, 
field from the other molecules. \cite{Ando94,Kubar2008,Valeev2006}
We thus first determine the 
effective atomic charges 
derived from the electrostatic potential (ESP) of an isolated alanine monomer,
in which both the structure and MOs are computed at the B3LYP/6-31G(d) level.
The results displayed in Figure S1 of the Supplementary materials
indicate that they are consistent with the AMBER 94 force-field \cite{Cornell1995}.

We now compute the basis MOs of each monomer in Figure 4 
under the classical field from the
ESP charges of others. 
The KS matrix of donor-acceptor pairs are similarly computed under the classical
field of remaining molecules.
This procedure is denoted by \lq Dimer in MM\rq.
The ordinary FOA without the MM field is denoted by \lq Dimer in Vac(uum)\rq.
These methods are summarized in Table 1 with respect to the orbital relaxation
effect in the basis MOs and the KS matrix.

\begin{table*}
Table 1.
Orbital relaxation in different computational schemes.
\\[1em]
\setlength{\tabcolsep}{1.5em}
\begin{tabular}{lllllll}
\hline \hline
          & {FMO-LCMO} &  Full FOA & \multicolumn{2}{c}{Dimer in MM} & \multicolumn{2}{c}{Dimer in Vac} \\
\cline{4-5}
\cline{6-7}
Orbital relaxation     &             &           &  FOA            &  GMH$^{c}$ &  FOA   &  GMH   \\
\hline
Basis MOs              &  Yes        &  No       &  MM$^{a}$       &  ---       &  No    &  ---   \\
KS matrix              &  Yes        &  Yes      &  MM+dimer$^{b}$ &  MM+dimer  &  Dimer &  Dimer \\
\hline \hline
\end{tabular}
\\[1em]
{}$^{a}$ The orbital relaxation is included via the MM field of point charges.
\\
{}$^{b}$ The orbital relaxation is included in the dimer calculation.
\\
{}$^{c}$ 
GMH computes the transfer integral from the dipole matrix and does not involve
the \lq basis MOs\rq.
\end{table*}

\begin{figure}
\begin{center}
\includegraphics[width=0.35\textwidth]{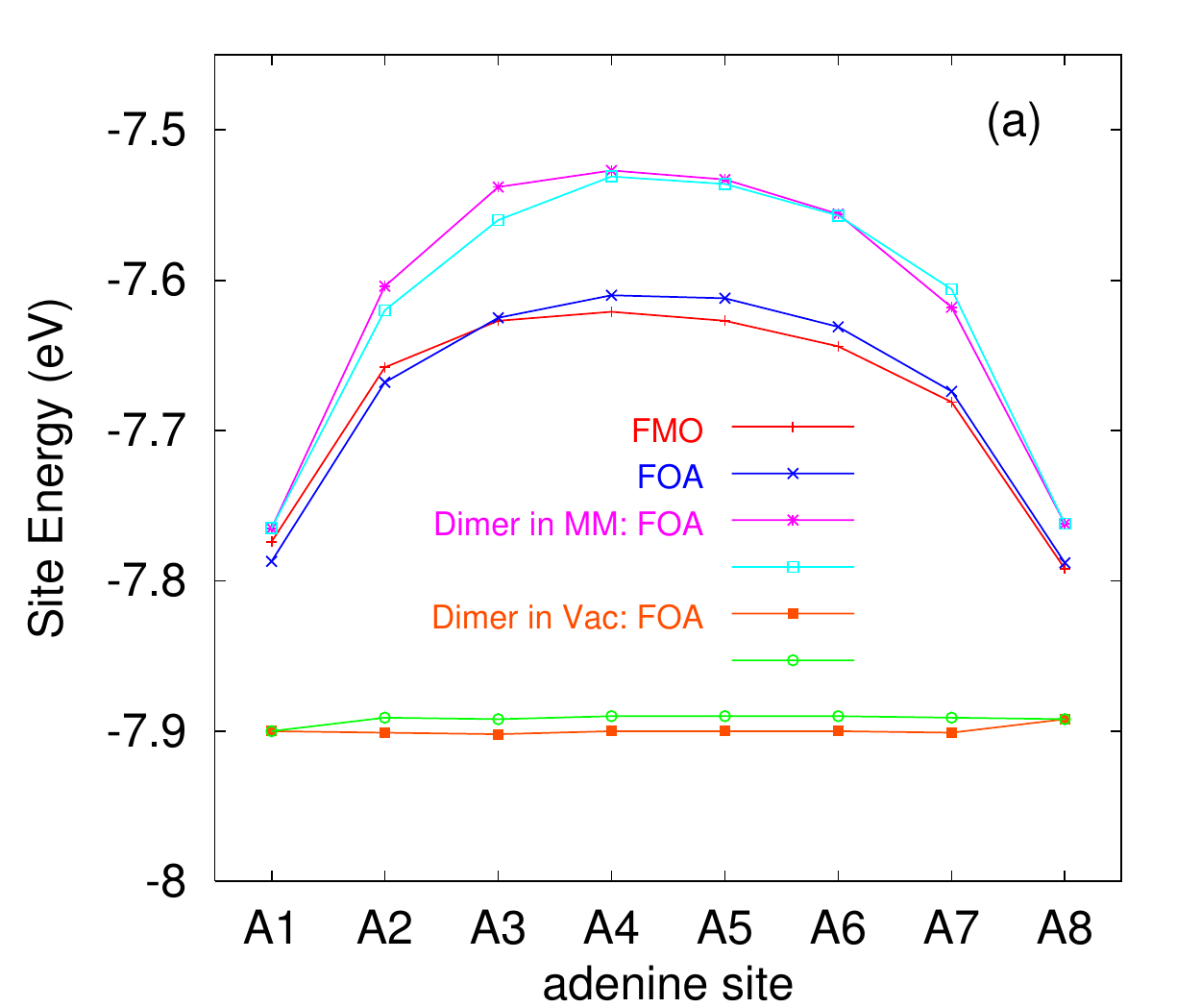}
\includegraphics[width=0.35\textwidth]{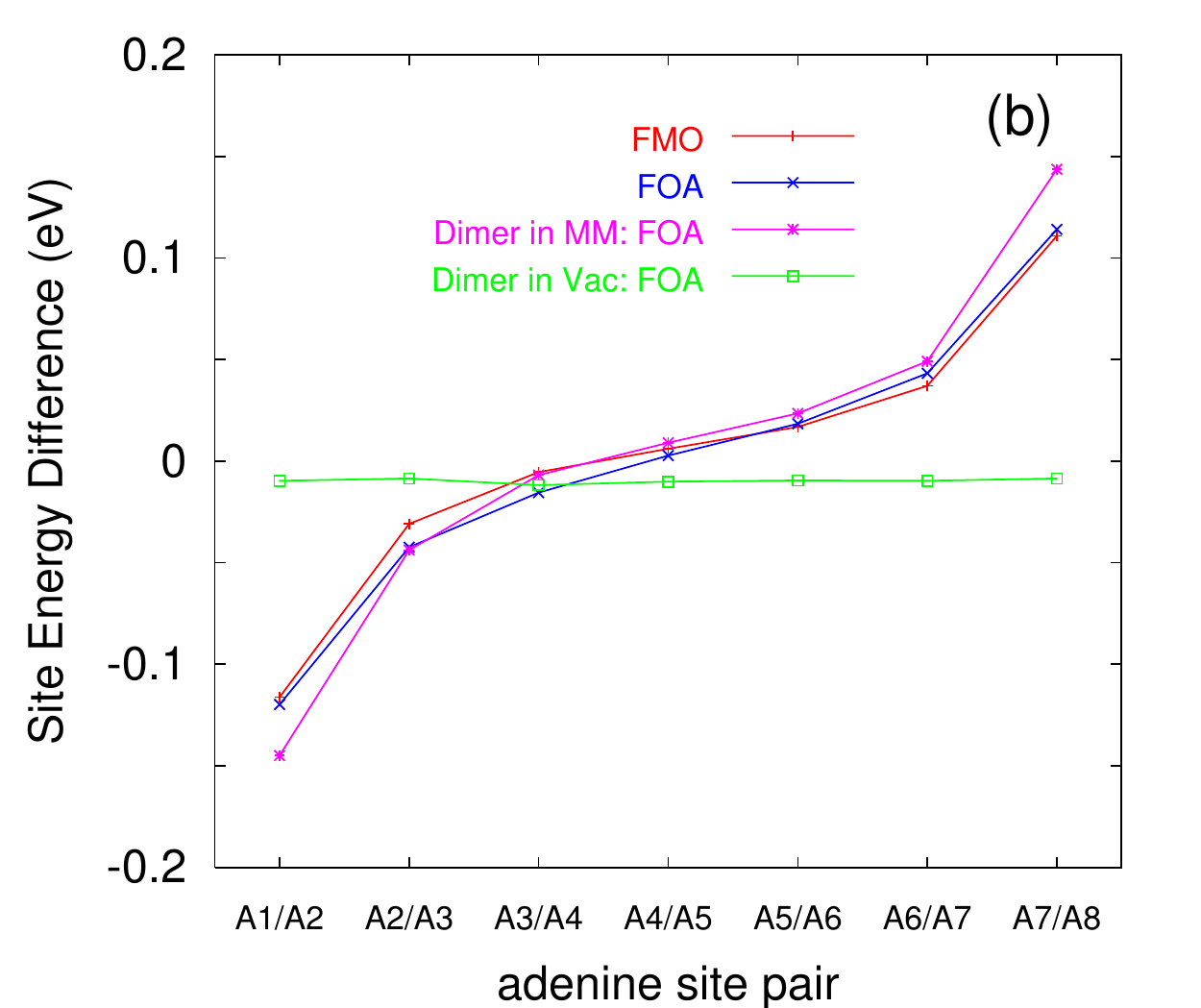}
\end{center}
\caption{
Site energies and their differences in the adenine octomer
with 6-31G(d) basis-set.
}
\end{figure}

Figure 5 plots the site energies 
of each adenine molecule and their differences $\Delta E_{ij}$ in the adjacent pairs.
The numerical values are listed in Table S5 of the Supplementary materials.
Figure 5a indicates that the site energies by the \lq Dimer in Vac\rq\ method
are almost independent of the position, 
naturally because the pairs are treated independently.
We also see that the absolute site energies are notably overestimated in the isolated pairs.
Contrastingly, 
the maximum of
the site energy near the middle of the chain A4-A5
is qualitatively captured by 
the \lq Dimer in MM\rq\ method, 
although the absolute energies are generally underestimated by approximately 0.1 eV.
Finally, we note that the site energies by the FMO-LCMO and the full FOA are consistent,
while the computational cost is notably reduced in the former.
Figure 5b indicates that the cancelation of errors in the site energies 
from the \lq Dimer in MM\rq\ method
results in correct profile of the site energy difference $\Delta E_{ij}$,
particularly in the region excluding the pairs at both ends of the chain.

\begin{figure}
\begin{center}
\includegraphics[width=0.35\textwidth]{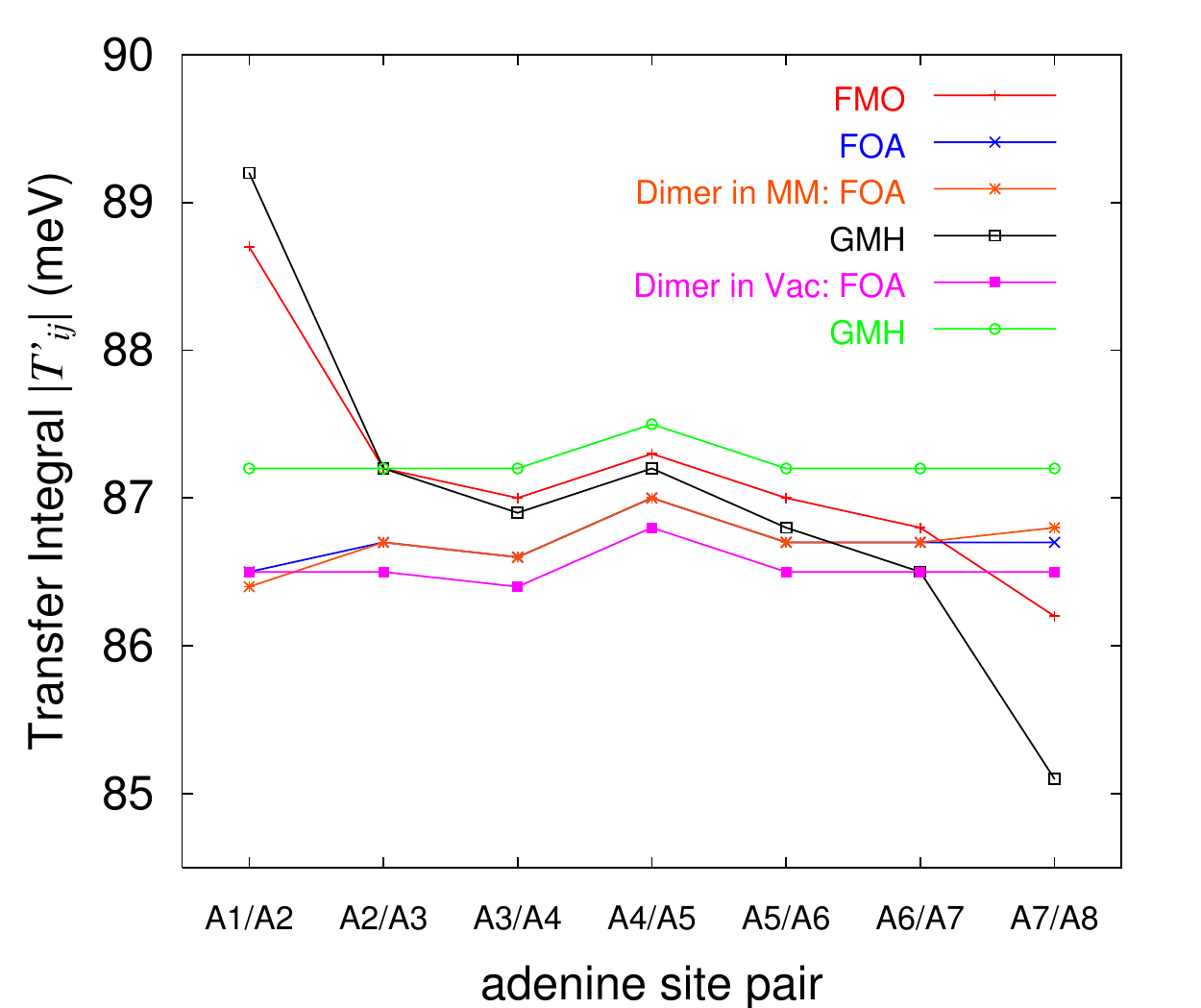}
\end{center}
\caption{
Transfer integrals between the adenine octomer
with 6-31G(d) basis-set.
}
\end{figure}

Figure 6 plots the transfer integrals $T'_{ij}$ between the HOMOs of adjacent adenine molecules.
The numerical values are listed in Table S6 of the Supplementary materials.
We first note that the results obtained from full FOA and 
FOA with \lq Dimer in MM\rq\ coincide well, 
with the exceptions in the A1/A2 and A7/A8 pairs at both ends.
To these two curves, that of FOA with \lq Dimer in Vac\rq\ is nearly parallel.
The same applies to the GMH with \lq Dimer in Vac\rq,
although the quantitative values are overestimated.
The profile of FMO-LCMO is overall different,
which is similar to that of
GMH with \lq Dimer in MM\rq.
The differences are most significant at both ends, A1/A2 and A7/A8,
which may be comprehended in terms of the balance between the 
inclusion of orbital relaxation effects in the basis MOs and in the KS matrix,
as presented in Table 1.

\subsection{Dependence on the basis-set}
\label{subsec:basisEffect}

\begin{figure*}
\begin{center}
\includegraphics[width=0.35\textwidth]{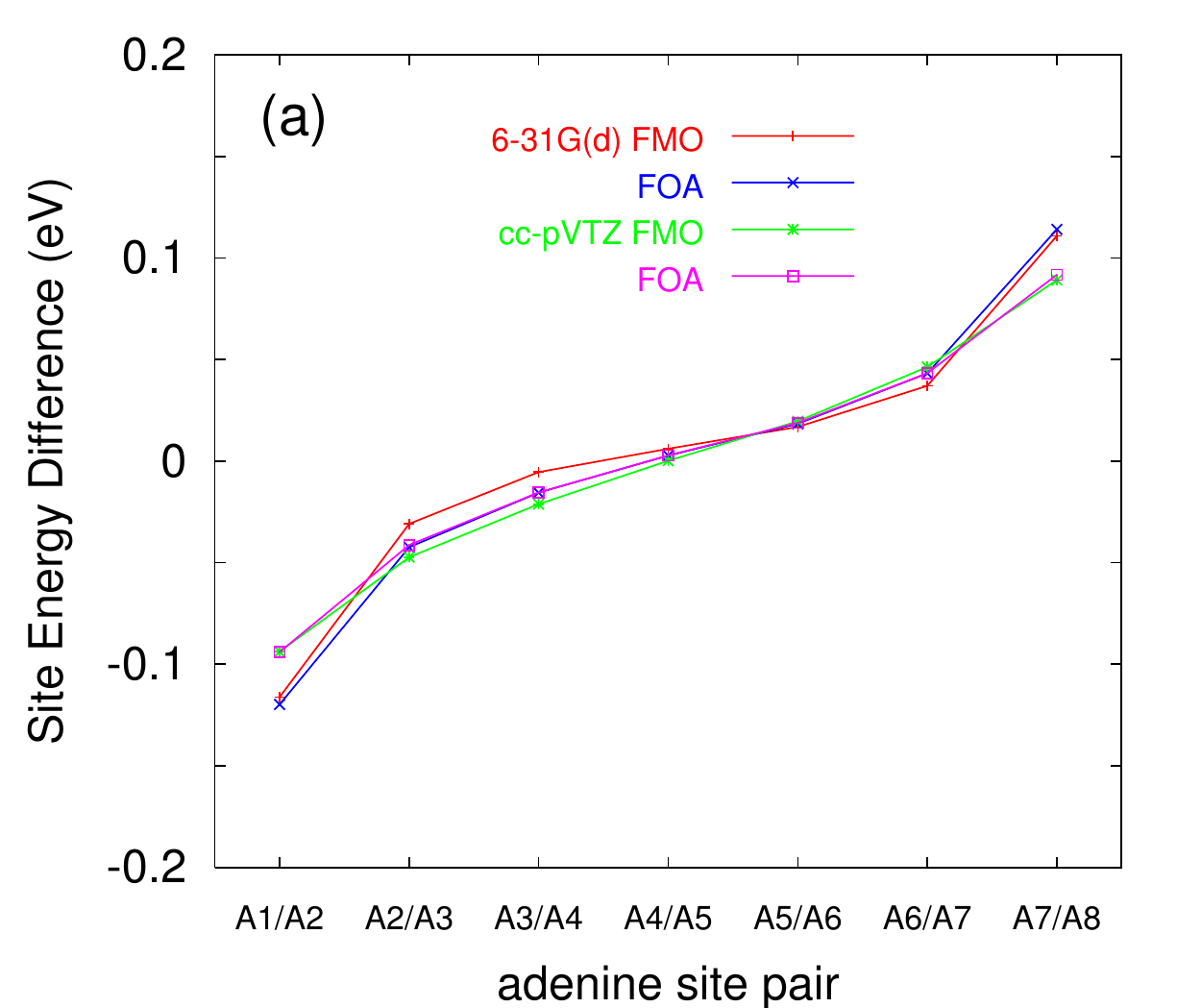}
\includegraphics[width=0.35\textwidth]{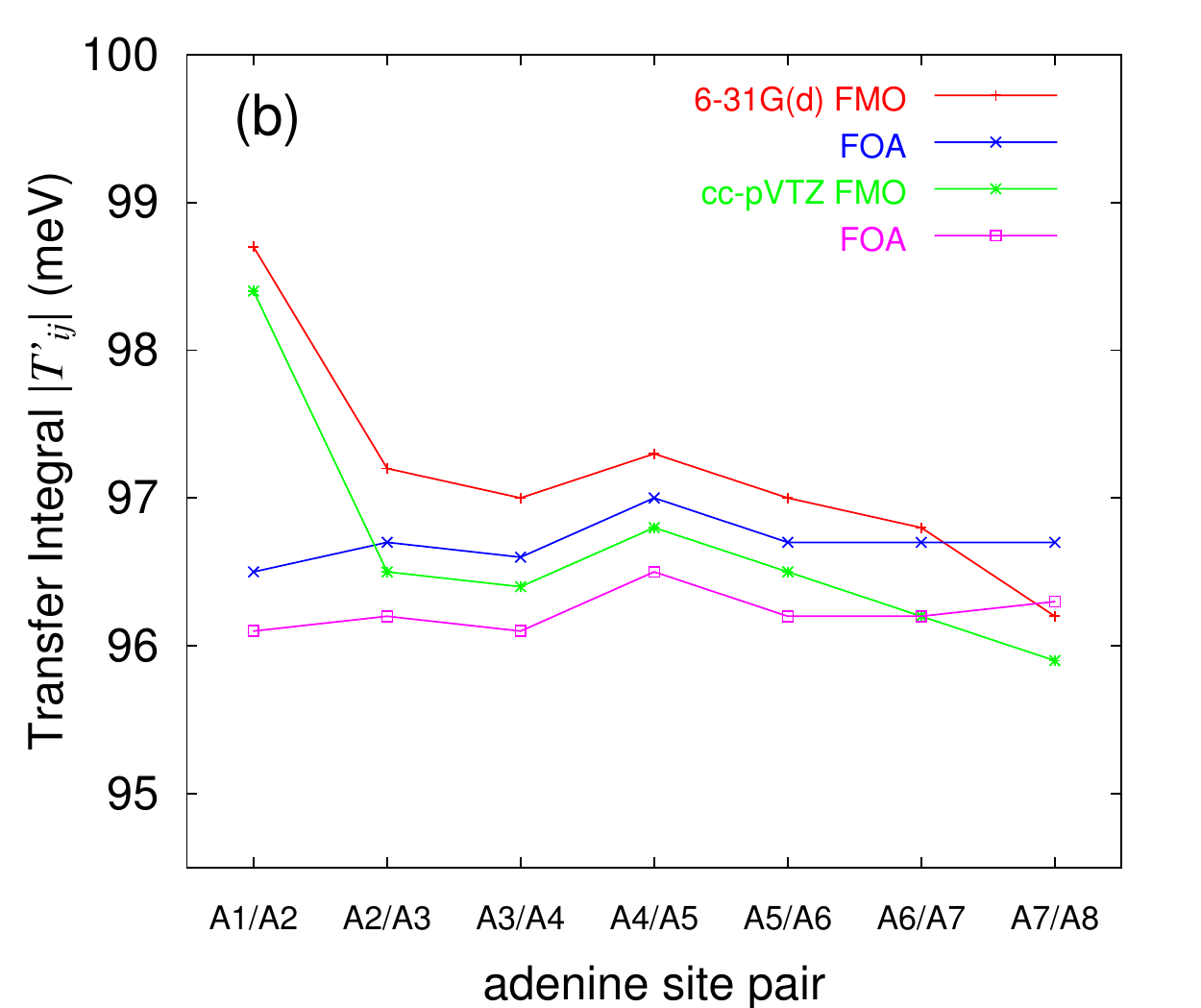}
\end{center}
\caption{
Comparison between 6-31G(d) and cc-pVTZ basis-sets for the
(a) site energy differences and (b) transfer integrals in the adenine octomer.
In (b), the values from 6-31G(d) basis-set 
were shifted by +10 meV for better view (see Figure 6).
}
\end{figure*}

Finally, we confirm that the pictures obtained in the previous sections
are basically independent of the size of the basis-set.
Figure 7 compares between 6-31G(d) and cc-pVTZ basis-sets for the 
site energy differences and the transfer integrals.
The numerical values are listed in Table S7 of the Supplementary materials.
Note that the values from 6-31G(d) are shifted by +10 meV in Figure 7b
for better view.
Except this quantitative shift, the overall qualitative behavior is mostly
identical between the two basis-sets.
Therefore, the use of 6-31G(d) can be recommended
for major part of studies, 
with additional quantitative corrections 
with larger basis-sets,
for instance, 
at selected key configurations
according to specific necessities.

\section{Conclusion}
\label{sec:conclusion}

A scheme for computing CT energies with a combination of the FMO-LCMO method
and the NET RS DFT 
was proposed, assessed on the benchmark systems of imidazole and furane
homo-dimer cations, 
and applied to hole transfers in DNA nucleobases.
The NET RS
functional optimized for intramolecular site energies
was found to give the intermolecular CT integrals 
notably close
to the MRCI+Q and NEVPT2 references.
In the applications to nucleaobases,
the FMO-LCMO gave the CT energies 
comparable to those with the full FOA 
at lower computational cost:
the efficiency will become more prominent as the system size increases.
The use of classical point charges to remedy the lack of orbital relaxation
in the dimer FOA calculations gave reasonable results
due to the cancelation of errors
in the site energy differences and transfer integrals.
Further details in the computational schemes were investigated in terms of
the orbital relaxation effect.

We have chosen to examine the hole transfer integrals of DNA nucleobases 
partly because of their own importance 
and partly in an aim to compare with the previous works.
However, the proposed combination of FMO-LCMO and NET RS DFT 
will be more effective 
in broad cases where the orbital relaxation and the exchange interactions are more significant.
For instance, when interactions with solvent and counter ions are involved,
the classical point-charge approximation will be less appropriate, particularly 
when their configurations thermally fluctuate.
This will come out crucial 
when positive classical charges
without the short-range exchange repulsion
overly attract the electrons.
Studies on such cases will be reported in due course.

\section*{Acknowledgments}

The authors acknowledge support from KAKENHI Nos. 20108017 (``$\pi$-space'') and 22550012.
H. K.-N. also acknowledges support from Collaborative Research Program for Young Scientists
of ACCMS and IIMC, Kyoto University.

\onecolumngrid

\newpage

\begin{table}
Table S1.
Transfer integrals $T'$ (in meV) in furane and imidazole dimer cations.
\\[0.8em]
\setlength{\tabcolsep}{0.5em}
\begin{tabular}{lcccccc}
\hline \hline
                & \multicolumn{2}{c}{LC-BLYP} & \multicolumn{2}{c}{B3LYP} & MRCI+Q$^a$ & NEVPT2$^a$ \\
\cline{2-3}
\cline{4-5}
Distance (\AA)  &  FMO2-LCMO & FOA & FMO2-LCMO & FOA & & \\
\hline
Furane & & & & & & \\
  3.5 &  425.9 &  424.9 &  361.2 &  360.5 &  440.3 &  412.7 \\
  4.0 &  213.8 &  213.6 &  174.6 &  174.5 &  214.9 &  200.8 \\
  4.5 &  104.4 &  104.6 &   83.2 &   83.3 &  101.8 &   97.4 \\
  5.0 &   49.3 &   49.5 &   39.2 &   39.4 &   46.0 &   48.9 \\
Imidazole & & & & & & \\
  3.5 &  419.9 &  418.5 &  355.6 &  354.6 &  411.6 & --- \\
  4.0 &  210.9 &  210.5 &  172.1 &  171.9 &  202.8 & --- \\
  4.5 &  103.2 &  103.3 &   82.2 &   82.3 &   99.1 & --- \\
  5.0 &   48.9 &   49.1 &   38.9 &   39.1 &   49.7 & --- \\
\hline\hline
\end{tabular}
\\[1em]
${}^{a}$ Ref. 37. 
\vspace*{1.5em}
\end{table}

\begin{table}
Table S2.
$-\varepsilon_{\rm HOMO-1}$ and $-\varepsilon_{\rm HOMO}$ (in eV) with LC-BLYP functional.
\\[0.8em]
\setlength{\tabcolsep}{1em}
\begin{tabular}{lcccc}
\hline \hline
  & \multicolumn{2}{c}{$\mu = 0.29$} & optimal $\mu$ $^{a}$ & expr. $^{b}$ \\
  \cline{2-3}
  & 6-31G(d) & cc-pVTZ & cc-pVTZ &  \\
\hline
Guanine  & 8.97 / 7.47 & 9.40 / 7.84 & 9.29 / 7.78 & 9.9  / 8.0-8.3 \\
Adenine  & 8.89 / 7.90 & 9.25 / 8.23 & 9.21 / 8.21 & 9.45 / 8.3-8.5 \\
Cytosine & 8.77 / 8.21 & 9.23 / 8.63 & 9.37 / 8.73 & 9.5  / 8.8-9.0 \\
Thymine  & 9.32 / 8.57 & 9.76 / 8.93 & 9.71 / 8.90 & 10.  / 9.0-9.2 \\
\hline \hline
\end{tabular}
\\[1em]
${}^{a}$ Ref. 9. 
\hspace*{0.8em}
${}^{b}$ Ref. 41. 
\vspace*{1.5em}
\end{table}

\begin{table}
Table S3.
Diabatic/adiabatic energy gaps (in eV) between nucleobase pairs with 6-31G(d) basis-set.
\\[0.8em]
\setlength{\tabcolsep}{0.6em}
\begin{tabular}{lccccc}
\hline \hline
          & \multicolumn{2}{c}{FMO2-LCMO} & \multicolumn{2}{c}{FOA} &  CASPT2$^{a}$ \\
\cline{2-3}
\cline{4-5}

base pair &  LC-BLYP &  B3LYP             &  LC-BLYP &  B3LYP  &   \\
\hline
GG    &  0.4074 / 0.4394  &  0.4071 / 0.4308  &  0.4081 / 0.4421  &  0.4078 / 0.4333 &  0.392  \\
GA    &  0.5629 / 0.5887  &  0.5287 / 0.5489  &  0.5646 / 0.5881  &  0.5312 / 0.5495 &  0.560  \\
AG    &  0.2588 / 0.2656  &  0.2202 / 0.2263  &  0.2598 / 0.2686  &  0.2214 / 0.2294 &  0.340  \\
GT    &  1.079  / 1.104   &  1.038  / 1.057   &  1.081  / 1.108   &  1.040  / 1.061  &  1.175  \\
TG    &  0.6782 / 0.6989  &  0.6309 / 0.6464  &  0.6795 / 0.7009  &  0.6325 / 0.6486 &  0.797  \\
\hline \hline
\end{tabular}
\\[0.8em]
${}^{a}$ Ref. 34. 
Adiabatic energy gap. 
\vspace*{1.5em}
\end{table}

\begin{table}
Table S4.
Transfer integrals $T'$ (in eV) between nucleobase pairs with 6-31G(d) basis set.
\\[0.8em]
\setlength{\tabcolsep}{0.5em}
\begin{tabular}{llllllll}
\hline \hline
          & \multicolumn{2}{c}{FMO2-LCMO} & \multicolumn{2}{c}{FOA} & \multicolumn{3}{c}{GMH} \\
\cline{2-3}
\cline{4-5}
\cline{6-8}

base pair &  LC-BLYP &  B3LYP             &  LC-BLYP &  B3LYP        &  LC-BLYP & B3LYP & CASPT2$^{a}$ \\
\hline
GG   & 0.08235   & 0.07052   & 0.08498   & 0.07326   & 0.08124   & 0.06981   & 0.051   \\
GA   & 0.08621   & 0.07372   & 0.08224   & 0.07035   & 0.08483   & 0.07264   & 0.036   \\
AG   & 0.02975   & 0.02614   & 0.03404   & 0.02996   & 0.02926   & 0.02589   & 0.044   \\
GT   & 0.1175    & 0.1006    & 0.1225    & 0.1055    & 0.1177    & 0.1012    & 0.081   \\
TG   & 0.08441   & 0.07044   & 0.08592   & 0.07181   & 0.08378   & 0.06971   & 0.061   \\
\hline \hline
\end{tabular}
\\[0.8em]
${}^{a}$ Ref. 34. 
\end{table}

\begin{figure}[t]
\begin{center}
\includegraphics[width=0.4\textwidth]{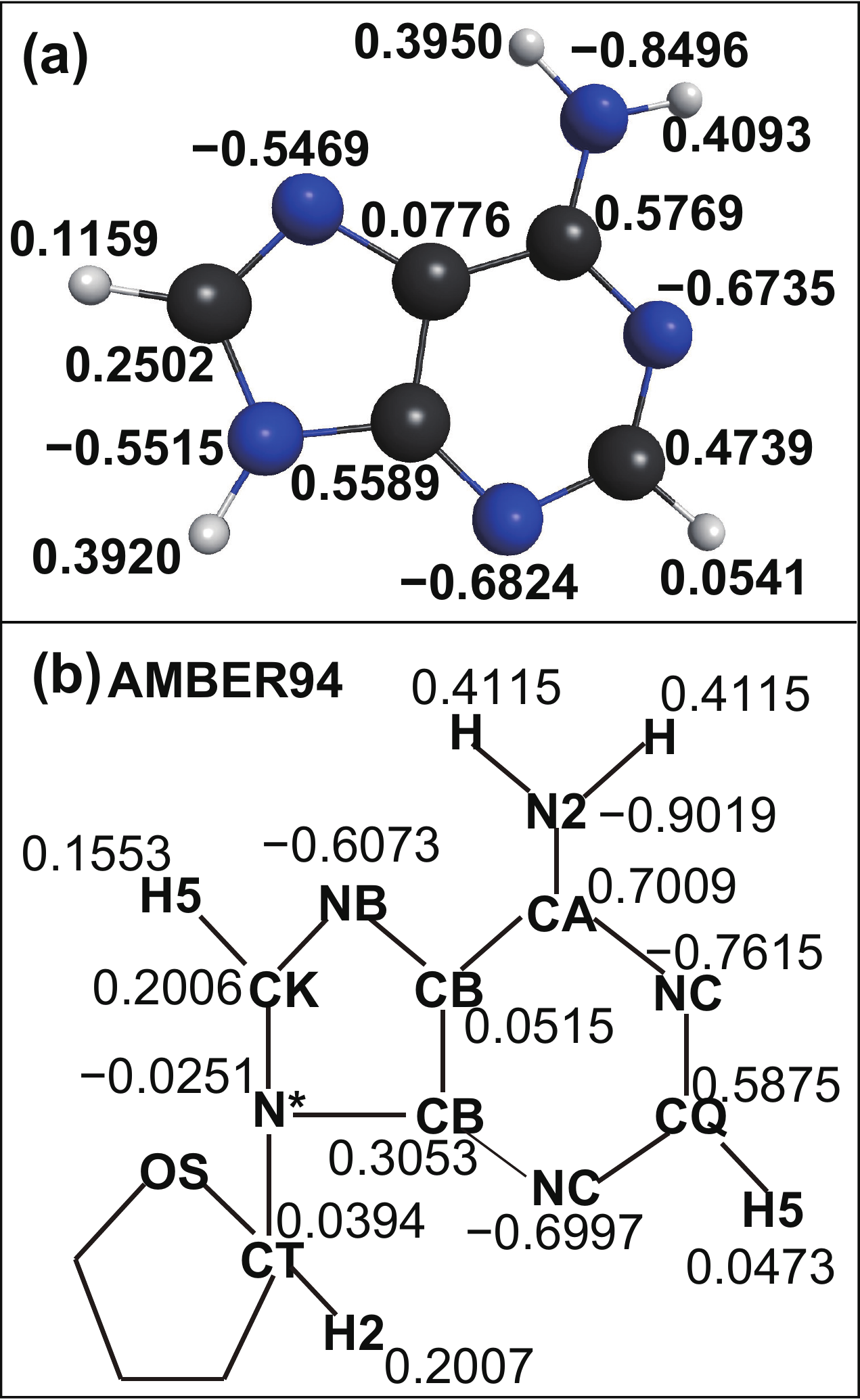}
\\
FIG. S1:
ESP charges on adenine.
\end{center}
\end{figure}

\begin{table}
Table S5.
Site energies (in eV) of $\pi$-stack eight adenine molecules with 6-31G(d) basis set.
\\[0.8em]
\setlength{\tabcolsep}{1em}
\begin{tabular}{lllll}
\hline \hline
          & {FMO2-LCMO} & \multicolumn{3}{c}{FOA} \\
\cline{3-5}
site      &             &  full   &  {Dimer in MM} &  {Dimer in Vac} \\
\hline
A1 &  -7.774  & -7.787  & -7.765            & -7.900           \\
A2 &  -7.658  & -7.668  & -7.604  / -7.620  & -7.901  / -7.891 \\
A3 &  -7.627  & -7.625  & -7.538  / -7.560  & -7.902  / -7.892 \\
A4 &  -7.621  & -7.610  & -7.527  / -7.531  & -7.900  / -7.890 \\
A5 &  -7.627  & -7.612  & -7.533  / -7.536  & -7.900  / -7.890 \\
A6 &  -7.644  & -7.631  & -7.556  / -7.557  & -7.900  / -7.890 \\
A7 &  -7.681  & -7.674  & -7.618  / -7.606  & -7.901  / -7.891 \\
A8 &  -7.792  & -7.788  &  \hspace*{2.7em} -7.762  & \hspace*{2.7em} -7.892 \\
\hline \hline
\end{tabular}
\end{table}

\begin{table}
Table S6.
Transfer integrals (in eV) of $\pi$-stack eight adenine molecules with 6-31G(d) basis set.
\\[0.8em]
\setlength{\tabcolsep}{1em}
\begin{tabular}{lllllll}
\hline \hline
          & {FMO2-LCMO} &  full FOA & \multicolumn{2}{c}{Dimer in MM} & \multicolumn{2}{c}{Dimer in Vac} \\
\cline{4-5}
\cline{6-7}
base pair &             &           &  FOA   &  GMH  &  FOA  &  GMH  \\
\hline
A1 / A2  &  0.0887  &  0.0865  &  0.0864  &  0.0892  &  0.0865  &  0.0872 \\
A2 / A3  &  0.0872  &  0.0867  &  0.0867  &  0.0872  &  0.0865  &  0.0872 \\
A3 / A4  &  0.0870  &  0.0866  &  0.0866  &  0.0869  &  0.0864  &  0.0872 \\
A4 / A5  &  0.0873  &  0.0870  &  0.0870  &  0.0872  &  0.0868  &  0.0875 \\
A5 / A6  &  0.0870  &  0.0867  &  0.0867  &  0.0868  &  0.0865  &  0.0872 \\
A6 / A7  &  0.0868  &  0.0867  &  0.0867  &  0.0865  &  0.0865  &  0.0872 \\
A7 / A8  &  0.0862  &  0.0867  &  0.0868  &  0.0851  &  0.0865  &  0.0872 \\
\hline \hline
\end{tabular}
\end{table}

\begin{table}
Table S7.
Site energy differences and transfer integrals (in eV) of $\pi$-stack eight adenine molecules 
with cc-pVTZ basis set.
\\[0.8em]
\setlength{\tabcolsep}{1em}
\begin{tabular}{lccccc}
\hline \hline
          & \multicolumn{2}{c}{Site Energy Difference} & & \multicolumn{2}{c}{Transfer Integral} \\
\cline{2-3}
\cline{5-6}
base pair & FMO2-LCMO &  FOA   & & FMO2-LCMO &   FOA   \\
\hline
A1 / A2  & -0.0938  & -0.0940  & &  0.0984  &  0.0961  \\
A2 / A3  & -0.0473  & -0.0413  & &  0.0965  &  0.0962  \\
A3 / A4  & -0.0212  & -0.0154  & &  0.0964  &  0.0961  \\
A4 / A5  &  0.0002  &  0.0028  & &  0.0968  &  0.0965  \\
A5 / A6  &  0.0196  &  0.0188  & &  0.0965  &  0.0962  \\
A6 / A7  &  0.0463  &  0.0433  & &  0.0962  &  0.0962  \\
A7 / A8  &  0.0891  &  0.0917  & &  0.0959  &  0.0963  \\
\hline \hline
\end{tabular}
\end{table}


\begin{thebibliography}{10}
\expandafter\ifx\csname url\endcsname\relax
  \def\url#1{\texttt{#1}}\fi
\expandafter\ifx\csname urlprefix\endcsname\relax\def\urlprefix{URL }\fi
\expandafter\ifx\csname href\endcsname\relax
  \def\href#1#2{#2} \def\path#1{#1}\fi

\bibitem{Marcus1985}
R.~A. Marcus, N.~Sutin, Biochim. Biophys. Acta 811 (1985) 265.

\bibitem{Bixon1999}
M.~Bixon, J.~Jortner, Adv. Chem. Phys. 106 (1999) 35.

\bibitem{Coropceanu2007}
V.~Coropceanu, J.~Cornil, D.~A. da~Silva~Filho, Y.~Olivier, R.~Silbey, J.-L. Br{\'e}das, 
Chem. Rev. 107 (2007) 926.

\bibitem{Grozema2008}
F.~C. Grozema, L.~D.~A. Siebbeles, Int. Rev. Phys. Chem. 27 (2008) 87.

\bibitem{Pieniazek07}
P.~A. Pieniazek, S.~A. Amstein, S.~E. Bradforth, A.~I. Krylov, C.~D. Sherrill,
  J. Chem. Phys. 127 (2007) 164110.

\bibitem{Oberhofer12}
H.~Oberhofer, J.~Blumberger, Phys. Chem. Chem. Phys. 14 (2012) 13846.

\bibitem{Iikura2001}
H.~Iikura, T.~Tsuneda, T.~Yanai, K.~Hirao, J. Chem. Phys. 115 (2001) 3540.

\bibitem{Tawada2004}
Y.~Tawada, T.~Tsuneda, S.~Yanagisawa, T.~Yanai, K.~Hirao,
  J. Chem. Phys. 120 (2004) 8425.

\bibitem{Foster2012}
M.~E. Foster, B.~M. Wong, J. Chem. Theory Comput. 8 (2012) 2682.

\bibitem{Stein2009}
T.~Stein, L.~Kronik, R.~Baer, J. Chem. Phys. 131 (2009) 244119.

\bibitem{Newton80}
M.~D. Newton, Int. J. Quant. Chem. S14 (1980) 363.

\bibitem{Yang06}
C.~H. Yang, C.~P. Hsu, J. Chem. Phys. 124 (2006) 244507.

\bibitem{Nishioka11a}
H.~Nishioka, K.~Ando, Phys. Chem. Chem. Phys. 13 (2011) 7043.

\bibitem{Pacher88}
T.~Pacher, L.~S. Cederbaum, H.~K{\"o}ppel, J. Chem. Phys. 89 (1988) 7367.

\bibitem{Ando94}
K.~Ando, J. Chem. Phys. 101 (1994) 2850.

\bibitem{Subotnik2008}
J.~E. Subotnik, S.~Yeganeh, R.~J. Cave, M.~A. Ratner, 
  J. Chem. Phys. 129 (2008) 244101.

\bibitem{Troisi2001}
A.~Troisi, G.~Orlandi, Chem. Phys. Lett. 344 (2001) 509.

\bibitem{Senthilkumar2005}
K.~Senthilkumar, F.~C. Grozema, C.~F. Guerra, F.~M. Bickelhaupt, F.~D. Lewis,
  Y.~A. Berlin, M.~A. Ratner, L.~D.~A. Siebbeles, 
  J. Am. Chem. Soc. 127 (2005) 14894.

\bibitem{Kubar2008}
T.~Kuba\v{r}, P.~B. Woiczikowski, G.~Cuniberti, M.~Elstner, 
  J. Phys. Chem. B 112 (2008) 7937.

\bibitem{QinWu06}
Q.~Wu, T.~Van Voorhis, J. Chem. Phys. 125 (2006) 164105.

\bibitem{delaLande10}
A.~de~la Lande, D.~R. Salahub, J. Mol. Struct. (Theochem) 943 (2010) 115.

\bibitem{Pavanello13}
M.~Pavanello, T.~Van Voorhis, L.~Visscher, J.~Neugebauer,
  J. Chem. Phys. 138 (2013) 054101.

\bibitem{Kitaura1999}
K.~Kitaura, E.~Ikeo, T.~Asada, T.~Nakano, M.~Uebayasi, 
  Chem. Phys. Lett. 313 (1999) 701.

\bibitem{Nakano2002}
T.~Nakano, T.~Kaminuma, T.~Sato, K.~Fukuzawa, Y.~Akiyama, M.~Uebayasi, K.~Kitaura,
  Chem. Phys. Lett. 351 (2002) 475.

\bibitem{Fedorov2007}
D.~G. Fedorov, K.~Kitaura, J. Phys. Chem. A 111 (2007) 6904.

\bibitem{Tanaka2014}
S.~Tanaka, Y.~Mochizuki, Y.~Komeiji, Y.~Okiyama, K.~Fukuzawa,
  Phys. Chem. Chem. Phys. 16 (2014) 10310.

\bibitem{Tsuneyuki2009}
S.~Tsuneyuki, T.~Kobori, K.~Akagi, K.~Sodeyama, K.~Terakura, H.~Fukuyama,
  Chem. Phys. Lett. 476 (2009) 104.

\bibitem{Kobori2013}
T.~Kobori, K.~Sodeyama, T.~Otsuka, Y.~Tateyama, S.~Tsuneyuki, 
  J. Chem. Phys. 139 (2013) 094113.

\bibitem{Nishioka2011}
H.~Nishioka, K.~Ando, J. Chem. Phys. 134 (2011) 204109.

\bibitem{Kitoh-Nishioka2012}
H.~Kitoh-Nishioka, K.~Ando, J. Phys. Chem. B 116 (2012) 12933.

\bibitem{Lowdin1950}
P.-O. L{\"o}wdin, J. Chem. Phys. 18 (1950) 365.

\bibitem{Cave97}
R.~J. Cave, M.~D. Newton, J. Chem. Phys. 106 (1997) 9213.

\bibitem{Felix2011}
M.~F{\'e}lix, A.~A. Voityuk, Int. J. Quant. Chem. 111 (2011) 191.

\bibitem{Blancafort2006}
L.~Blancafort, A.~A. Voityuk, J. Phys. Chem. A 110 (2006) 6426.

\bibitem{Valeev2006}
E.~Valeev, V.~Coropceanu, D.~A. da{\ }Silva{\ }Filho, S.~Salman, J.-L. Br{\'e}das,
  J. Am. Chem. Soc. 128 (2006) 9882.

\bibitem{Schmidt1993}
M.~W. Schmidt, K.~K. Baldridge, J.~A. Boatz, S.~T. Elbert, M.~S. Gordon,
  J.~H. Jensen, S.~Koseki, N.~Matsunaga, K.~A. Nguyen, S.~Su, T.~L. Windus,
  M.~Dupuis, J.~A. Montgomery{\ }Jr, 
  J. Comput. Chem. 14 (1993) 1347.

\bibitem{Kubas2014}
A.~Kubas, F.~Hoffmann, A.~Heck, H.~Oberhofer, M.~Elstner, J.~Blumberger,
  J. Chem. Phys. 140 (2014) 104105.

\bibitem{Becke1988}
A.~D. Becke, Phys. Rev. A 38 (1988) 3098.

\bibitem{Lee1988}
C.~Lee, W.~Yang, R.~G. Parr, Phys. Rev. B 37 (1988) 785.

\bibitem{Becke1993}
A.~D. Becke, J. Chem. Phys. 98 (1993) 5648.

\bibitem{Faber2011}
C.~Faber, C.~Attaccalite, V.~Olevano, E.~Runge, X.~Blase, 
  Phys. Rev. B 83 (2011) 115123.

\bibitem{Zheng2009}
G.~Zheng, X.-J. Lu, W.~K. Olson, 
  Nucleic Acids Res. 37~(suppl 2) (2009) W240.

\bibitem{Cornell1995}
W.~D. Cornell, P.~Cieplak, C.~I. Bayly, I.~R. Gould, K.~M. Merz, D.~M.
  Ferguson, D.~C. Spellmeyer, T.~Fox, J.~W. Caldwell, P.~A. Kollman, 
  J. Am. Chem. Soc. 117 (1995) 5179.

\end{thebibliography}
\end{document}